\documentclass[aps,prb,twocolumn,superscriptaddress]{revtex4-2}
\usepackage{graphicx}
\usepackage{bm}
\usepackage{amsmath}
\usepackage{amssymb}
\usepackage{amsfonts}
\usepackage{euscript}
\usepackage{verbatim}
\usepackage{setspace}
\usepackage{xcolor}
\usepackage{amsfonts}
\usepackage{braket}
\usepackage{qcircuit}
\usepackage[unicode=true]{hyperref}

\begin{document}
\title{Playing quantum nonlocal games with six noisy qubits on the cloud}
\author{Meron Sheffer}
\affiliation{Department of Physics, Bar-Ilan University, Ramat Gan 5290002, Israel}
\affiliation{Center for Quantum Entanglement Science and Technology, Bar-Ilan University, Ramat Gan 5290002, Israel}
\author{Daniel Azses}
\affiliation{School of Physics and Astronomy, Tel Aviv University, Tel Aviv 6997801, Israel}
\author{Emanuele G. Dalla Torre}
\affiliation{Department of Physics, Bar-Ilan University, Ramat Gan 5290002, Israel}
\affiliation{Center for Quantum Entanglement Science and Technology, Bar-Ilan University, Ramat Gan 5290002, Israel}

\begin{abstract}

Nonlocal games are extensions of Bell inequalities, aimed at demonstrating quantum advantage. These games are well suited for noisy quantum computers because they only require the preparation of a shallow circuit, followed by the measurement of non-commuting observable. Here, we consider the minimal implementation of the nonlocal game proposed in Science 362, 308 (2018). We test this game by preparing a 6-qubit cluster state using quantum computers on the cloud by IBM, Ionq, and Honeywell. Our approach includes several levels of optimization, such as circuit identities and error mitigation and allows us to cross the classical threshold and demonstrate quantum advantage in one quantum computer. We conclude by introducing a different inequality that allows us to observe quantum advantage in less accurate quantum computers, at the expense of probing a larger number of circuits.
\end{abstract}

\maketitle

A fundamental result of quantum information is the experimental violation of the bounds of classical theories. Historically, this result was first achieved by the violation of Bell inequalities using a pair of entangled photons \footnote{See Refs.~\cite{clauser1969proposed,freedman1972experimental} for the original proposal and realization, and Ref.~\cite{brunner2014bell} for a review of following criticisms and loophole-free experiments.}. In the last decade, researchers were able to extend these violations to large number of degrees of freedom \cite{walther2005experimental,gao2010bell,yao2012observation,lanyon2014experimental}. Notable examples are the violation of a 8-qubit Bell inequality with entangled photons \cite{yao2012observation}, and a 14-qubit inequality with trapped ions \cite{lanyon2014experimental}. Violations of Bell inequalities can be equivalently described as achievements of quantum advantage in nonlocal games \footnote{See Ref.~\cite{khan2018quantum} for an introduction to quantum games and Ref.~\cite{silman2008relation} for the equivalence of Bell inequalities and nonlocal games}.

\begin{figure}[b]
	\centering
	\includegraphics[width=\linewidth]{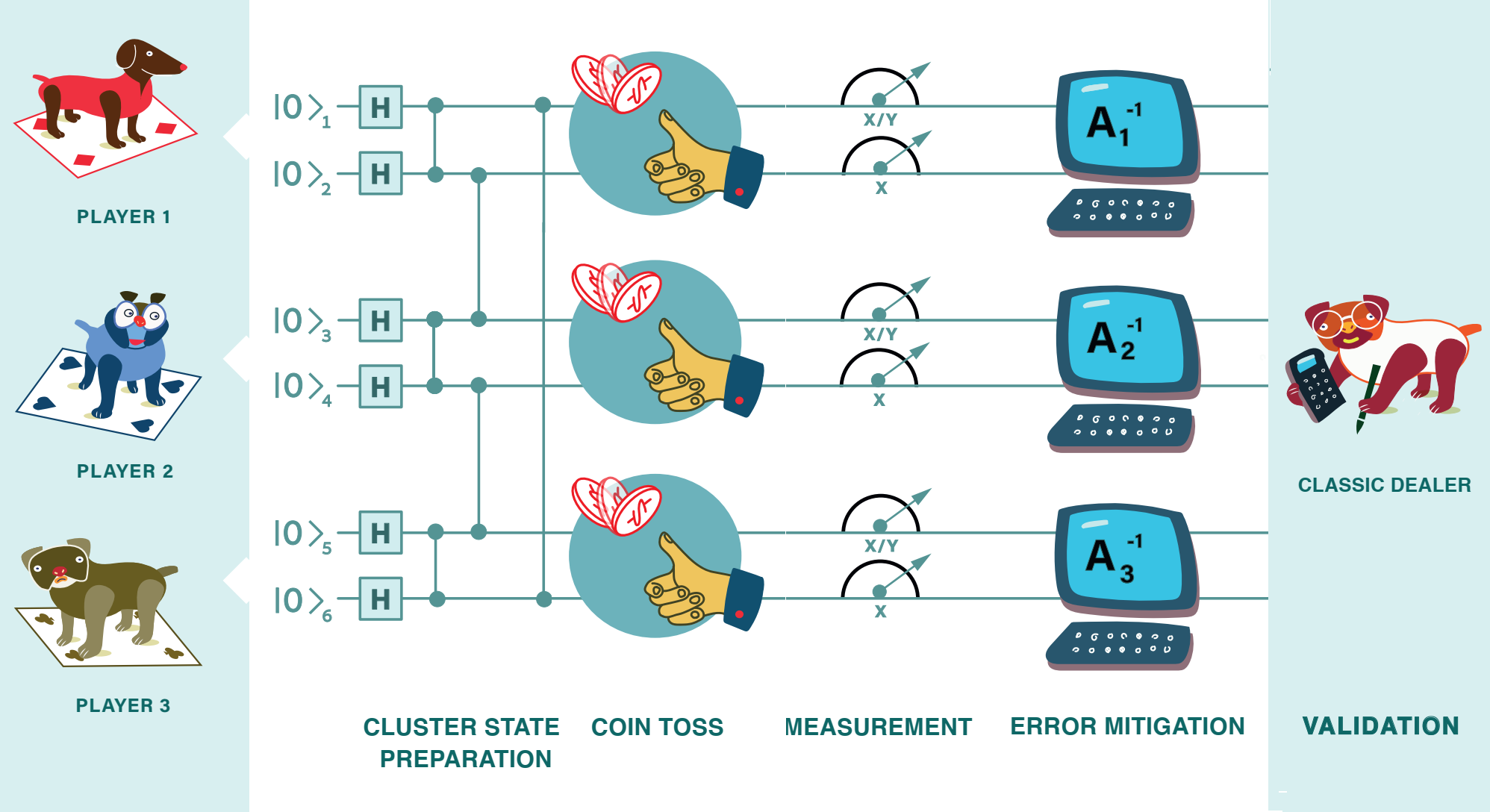}
	\caption{Schematic description of the optimal quantum strategy for the minimal realization of the triangle game of Ref.~\cite{bravyi2018quantum}. The game involves 3 players, each having access to 2 qubits. The qubits are initialized  in a one-dimensional cluster state with periodic boundary conditions. Next, each player tosses a coin, performs a quantum measurement according to the result of the coin, and applies an error-mitigation protocol to the result. The classical dealer collects the results of all the players and verifies if they satisfy the winning conditions associated with the results of the coins (see Table \ref{table1}). In the absence of noise, the quantum players satisfy all the conditions and win the game with a 100\% probability, as opposed to the optimal classical strategy, whose winning probability is $87.5$\%. }
	\label{fig:scheme}
\end{figure}

A notorious nonlocal game in which quantum mechanics shows an advantage over classical theories was introduced by Ref.~\cite{bravyi2018quantum}. This collaborative game, which we refer to as the {\it triangle game}, requires three or more players to independently assign binary values to a set of variables, aiming to satisfy some global winning conditions. A key aspect of the triangle game is that, for each instance of the game, the winning conditions are determined by three stochastic binary variables ({\it coins} or {\it inputs}), each tossed by a different player. The conditions are defined in a way that the best classical strategy can satisfy them only in 7 out of 8 cases, i.e. with an average probability of 87.5\%. In contrast, quantum players can win the games with 100\% probability by sharing a quantum  entangled state and performing independent measurements according to the result of their coin. See Fig.~\ref{fig:scheme} for the minimal implementation of the game, involving 3 players and 6 qubits.

The triangle game is built upon the quantum properties of a paradigmatic quantum information state, the one dimensional {\it cluster} state. This state is the building block of several quantum algorithm, such as measurement-based quantum teleportation and computation \cite{raussendorf2001one,else2012symmetry,else2012symmetry_long,choo2018measurement,azses2020identification,azses2020symmetry}. The cluster state is the common eigenstate of a set of commuting Hermitian operators with eigenvalue $\pm 1$, named {\it stabilizers}. For a $n$qubit cluster state with periodic boundary conditions, the stabilizers are $s_i = Z_{i-1}X_{i}Z_{i+1}$, where $X_i$ and $Z_i$ are Pauli operators and we use the notation $Z_{n+1}=Z_1$ and $Z_{0}=Z_{n}$. The cluster state is the only state satisfying $s_i|\psi\rangle=|\psi\rangle$ for all $i$, i.e. where a measurement of all products of $s_i$ always gives $+1$. Importantly, this state can be easily prepared on a quantum computer by applying Hadamard gates on all qubits, followed by control-Z gates on neighboring qubits \cite{choo2018measurement,azses2020identification}.

\begin{table}[t]
	\centering
	\begin{tabular}{|c |c |c| c|}
		\hline
		& $aaa$ & $bbb$ & $dea$\\
		\hline
		000 & $X_1 X_3 X_5 = 1$ & $X_2 X_4 X_6=1$ & \\
		\hline
		001 & & $X_2 X_4 X_6=1$ & \\
		\hline
		010 & & $X_2 X_4 X_6=1$ & \\
		\hline
		011 &  & $X_2 X_4 X_6=1$ & $X_1 Y_3 X_4 Y_5=-1$\\
		\hline
		100 &  & $X_2 X_4 X_6=1$ &\\
		\hline
		101 & & $X_2 X_4 X_6=1$ &  $Y_1 X_3 Y_5 X_6=-1$\\
		\hline
		110 & & $X_2 X_4 X_6=1$ & $Y_1 X_2 Y_3 X_5=-1$\\
		\hline
		111 & & $X_2 X_4 X_6=1$ & \\
		\hline
	\end{tabular}	
	\caption{Winning conditions of the triangle game, expressed in terms of products of Pauli operators \cite{daniel2021quantum}. The first column refers to the result of the coin tossing, and the other columns provide the relevant winning conditions. Empty cells refer to conditions that do not apply. The player win the game if the relevant winning condition(s) is (are) satisfied.}
	\label{table1}
\end{table}

The relation between the stabilizers of the cluster state and the winning conditions of the triangle game has been recently shown in Ref.~\cite{daniel2021quantum}. Following their notation, the winning conditions of the triangle game can be classified in three categories, respectively denoted by $aaa$, $bbb$ and $dea$. Each condition requires that a product of Pauli operators equals to $+1$ or $-1$, see Table \ref{table1}. Quantum players can satisfy all the conditions by sharing a cluster state: the left-hand-side of each condition corresponds to the product of stabilizers, as can be see from the identities $ X_{i}X_{i+2}X_{i+4} = s_i s_{i+2} s_{i+4}$ and $Y_{i}X_{i+1}Y_{i+2}X_{i+4}= -s_{i} s_{i+1} s_{i+2} s_{i+4}$. Note that the latter identity includes a minus sign, generated by the anticommutation relation of Pauli operators.
Hence, all the conditions are always satisfied if the $i$th player measures $(X_{2i-1},X_{2i})$ when her/his coin is 0, and $(Y_{2i-1},X_{2i})$ when her/his coin is 1. Using this strategy, the quantum players win the triangle game with $100\%$  probability.

In contrast, if we assume a local hidden variable (LHV) model where each Pauli operator acquires a well-defined value, the winning conditions of Table \ref{table1} cannot be simultaneously satisfied. The best classical strategy is obtained by the classical strategy $(X_{2i-1}, X_{2i},Y_{2i})=(-1,1,1)$. This choice satisfies all but the $aaa$ condition and wins in $7/8=87.5\%$ of the cases. Hence, for this game, the optimal classical-to-quantum ratio is $R=7/8=87.5\%$. For later reference, we also consider the case of players that play randomly: Because in 4 out of 8 games, only one condition applies, while in the remaining 4 two conditions apply, the average winning probability will be $(25\%+50\%)/2=37.5\%$.

\begin{table*}
	\renewcommand{\arraystretch}{1.5}
	\begin{tabular}{|c |c|c |c|c|}
		\hline
		Quantum Computer & Qubits & Cloud service & Programming Language & Price \\
		\hline
		\hline
		\begin{tabular}{c} 
		ibm\_melbourne 
		\\ ibm\_boeblingen \end{tabular} & 
		\begin{tabular}{c} 
		15 
		\\ 27 \end{tabular} 
		& IBM Quantum & QISKIT (Python) & 
		free \\
		\hline
		IonQ & 11 & Amazon AWS Braket & Amazon Braket SDK (Python)& 10\$\\
		\hline
		\begin{tabular}{c} 	Honeywell H0 \\ Honeywell H1 \end{tabular} & 
		\begin{tabular}{c}  6 \\ 10 \end{tabular}  
			 & Microsoft Azure Quantum & Qsharp & \begin{tabular}{c}  35\$ \\ -- \end{tabular} \\
		\hline
	\end{tabular}
	\caption{Quantum computers, cloud services, and programming language that were used in this project. The price refers to 1024 measurements (shots) of a single circuit. The Honeywell quantum computer H1 is not priced on a pay-per-use plan. The IBM quantum computers ibm\_melbourne and ibm\_boeblingen have been retired and are not currently on the cloud.
	}
	\label{table2}
\end{table*}

We now move to the experimental realization of the triangle game using quantum computers on the cloud. The quantum computers that we used and the relative costs are listed in Table \ref{table2}. We realize the game by preparing 8 circuits, one for each set of winning conditions in Table \ref{table1}. In each circuit we, first, prepare  the cluster state using $H$ and control-Z gates, and then apply to each qubit a $H$ or $H S^\dagger$ gate, depending on whether the player needs to measure the $X$ or $Y$ Pauli operator. We assign $+1$ or $-1$ to the 0 and 1 results of the quantum measurements, respectively, and check whether the winning conditions of Table \ref{table1} are satisfied. The total winning probability is set by the average result of this procedure on the 8 circuits. Our initial attempts to run the triangle game on the cloud lead to very low winning probabilities, close to the random strategy. To improve our winning probability, we introduced three optimization techniques. 

First, in the case of superconducting qubits, we used the connectivity of the quantum processors to select which physical qubits are used by the quantum computer. To achieve an optimal fidelity, we selected a set of 6 neighboring qubits connected with each other in a closed loop. This requirement prevented us from using some state-of-the-art quantum computers, such as the IBM Montreal and Rigetti Aspen, where the qubits follow octagonal structures, and no closed loop with 6 qubits is available. In principle, one can apply quantum gates between non-neighboring qubits using intermediate SWAP gates, but this choice leads to a significant deterioration of the result. Trapped-ions quantum computers have all-to-all couplings and, hence, are unaffected by this problem.

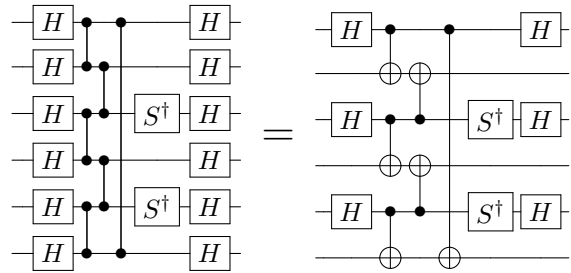
\begin{figure}[t]
	\centering
	\begin{tabular}{c c c}
		\begin{tabular}{c}	
			\Qcircuit @C=0.4em @R=.4em {		
				\\ & \qw &\gate{H} & \ctrl{1} &\qw&\ctrl{5}&\qw&\gate{H}&\qw 		
				\\ & \qw &\gate{H} & \ctrl{-1} &\ctrl{1}&\qw&\qw&\gate{H}&\qw 		
				\\ & \qw &\gate{H} & \ctrl{1} &\ctrl{-1}&\qw&\gate{S^\dagger}&\gate{H}&\qw 		
				\\ & \qw &\gate{H} & \ctrl{-1} &\ctrl{1}&\qw&\qw&\gate{H}&\qw 		
				\\ & \qw &\gate{H} & \ctrl{1} &\ctrl{-1}&\qw&\gate{S^\dagger}&\gate{H}&\qw 		
				\\ & \qw &\gate{H} & \ctrl{-1} &\qw&\ctrl{-5}&\qw&\gate{H}&\qw 		
		} \end{tabular}
		& \begin{tabular}{c}\\ \LARGE =\end{tabular}  &
		\begin{tabular}{c}	
			\Qcircuit @C=0.3em @R=.61em {	
				\\ & \qw &\gate{H} & \ctrl{1} &\qw&\ctrl{5}&\qw&\gate{H}&\qw 	
				\\ & \qw &\qw & \targ &\targ&\qw&\qw&\qw&\qw 	
				\\ & \qw &\gate{H} & \ctrl{1} &\ctrl{-1}&\qw&\gate{S^\dagger}&\gate{H}&\qw 	
				\\ & \qw &\qw & \targ &\targ&\qw&\qw&\qw&\qw 	
				\\ & \qw &\gate{H} & \ctrl{1} &\ctrl{-1}&\qw&\gate{S^\dagger}&\gate{H}&\qw 	
				\\ & \qw &\qw & \targ &\qw&\targ&\qw&\qw&\qw 	
			}
		\end{tabular}	
	\end{tabular}
	\caption{Two equivalent quantum circuits, used to measure $\langle X_1 X_2 Y_3 X_4 Y_5 X_6\rangle $. These circuits realize the optimal quantum strategy for the coin tossing result '011'. The right version of the circuit is optimized for running on noisy quantum computers.}
	\label{fig_quantum_circ}
\end{figure}

The second optimization consisted in  using circuit identities to reduce the number of gates. In particular, a control-Z gate is equivalent to a control-NOT gate, followed and succeeded by an $H$ gate on the controlled qubit. For some systems, such as the IonQ processor, this substitution is anyway necessary because the hardware supports control-NOT gates only and cannot process control-Z gates. We then use the $H^2=1$ identity to reduce the total number of $H$ gates in the circuit, see for example Fig.~\ref{fig_quantum_circ}, where this substitution allows us to reduce the number of $H$ gates from $12$ to $6$. 

The third optimization step aimed at fixing state-preparation and measurement (SPAM) errors, by post-processing the experimental outcomes. Specifically, we implemented the linear error mitigation protocol of Ref.~\cite{bravyi2021mitigating}. This method is based on the physical assumption that SPAM errors can be described as a stochastic process where the probability to obtain the outcome $i$ in the noisy circuit equals to the probability to obtain the outcome $j$ in the clean circuit, times a constant probability $(A)_{i,j}$. Under these assumptions, SPAM errors can be fixed in three steps: (i) estimate the matrix $A$ by performing calibration measurements of all possible outcomes of the circuit, (ii) compute the matrix $A^{-1}$ by numerically inverting $A$, (iii) multiply the histograms of the experimental results by $A^{-1}$.

The naive application of this method requires each player to know the output of all other players and, hence, is inconsistent with the rules of nonlocal quantum games, which forbid any classical communication among the players. To overcome this problem, we introduce a {\it local} mitigation process, where each player modifies the histograms of her/his own outcomes, independently from the results of the other players. Our approach consists in building an error mitigation matrix using the tensor-product form $A^{-1} = A_1^{-1}\times A_2^{-1} \times A_3^{-1}$, where $A_{i}^{-1}$ is a $2\times2$ matrix, acting on the qubits associated with the player $i$ only, namely $(2i-1,2i)$. In our experiment, we built the matrices $A_{i}^{-1}$ using 4 calibration circuits that prepared and measured, respectively, the states (00000), (010101), (101010), and (11111). For each pair of qubits we, then, used the calibration measurements to estimate the matrix $A_i$, and numerically computed $A_i^{-1}$. We then multiplied the 3 matrices to obtain $A^{-1}$ and used it to mitigate the errors in the experimental measurements.

\begin{figure}[b]
	\centering
	\includegraphics[trim=2.2cm 9.5cm 2.2cm 9.5cm, clip=true,width=\linewidth]{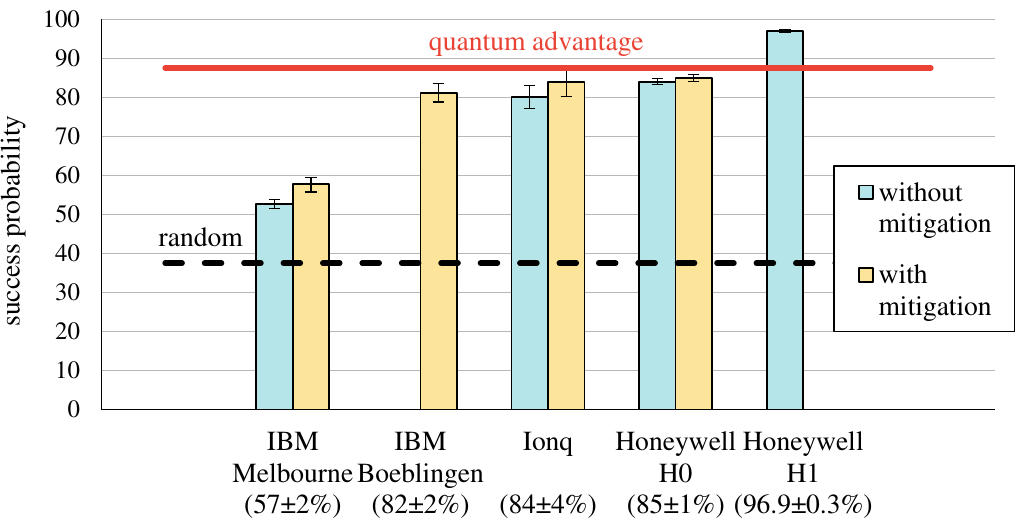}
	\caption{Winning probability of the triangle game by Ref.~\cite{bravyi2018quantum} on 5 quantum computers on the cloud. (The numbers in parenthesis are the error-mitigated results.)}
	\label{fig:result}
\end{figure}

Our results for the winning probability of the triangle game in 5 quantum computers, with and without error mitigation, are shown in Fig.~\ref{fig:result} \footnote{The codes used to generate the quantum circuits described in this article and the measurement results can be found online at \url{https://github.com/quantum-data/quantum-non-local}}. Each realization of the game involved 1024 measurements of 12 quantum circuits (4 circuits for the calibration and 8 circuits for the game). The  error-bars were obtained by repeating the full procedure (calibration and nonlocal game) several times. The results obtained with SPAM error mitigation are consistently better than those obtained without mitigation, with an average improvement of a few percent only. As a main result, we found that only one quantum computer, the Honeywell H1, had a winning probability larger than the classical threshold of 87.5\% and demonstrated a quantum advantage.

In what follows, we develop an alternative strategy that allows us to observe quantum advantage of the cluster state, in computers where the triangle game was unsuccessful. Our approach is based on Bell inequalities expressed in terms of products of sums of stabilizers. In the case of a cluster state with $n=6$ qubits, there are $2^n=64$ distinct products of stabilizers. The sum of all these products, which we denote by $S_{\rm all}$, has an expectation value of $64$. For classical LHV models, it was shown that $S_{\rm all}\leq28$ \cite{guhne2005bell}.  Following Ref.~\cite{silman2008relation}, we can translate this Bell inequality into a nonlocal game, where quantum players win with a 100\% probability, and classical players can, at most, with a probability of $(1+28/64)/2\approx 72\%$.
	
An interesting question is whether this probability can be further lowered by considering a different sum of products of stabilizers. This question was answered positively in Ref.~\cite{cabello2008mermin}, where it was shown that the optimal sum is given by
\begin{align}
S_{\rm optimal} = S_{\rm all} - 1 - \sum_{i=1}^6 s_i - \sum_{i=1}^2 s_{i}s_{i+2}s_{i+4}	
\label{eq:optimal}
\end{align}
The operator $S_{\rm optimal}$ corresponds to the sum of 55 products of stabilizers and, hence, its expectation value in the cluster state is $55$. In contrast, the maximal LHV value from all possible configurations of the Pauli operators is 19, leading to a nonlocal game with winning probability of $(1+19/55)/2 \approx 67\%$. This value is much smaller than the classical winning probability of the triangle game and closer to experimental reach. 

To experimentally demonstrate quantum advantage, we automatically generated 55 quantum circuits, one for each term of $S_{\rm optimal}$, and averaged their results. We realized this protocol on the IonQ quantum computer, without any optimization step, and obtained an expectation value of $\langle S_{\rm optimal}\rangle = 41\pm 0.5$, corresponding to a winning probability of $[1+(41\pm0.5)/55]/2 = 87\pm 1\%$, which is significantly above the classical winning probability and demonstrates a quantum advantage. While we have not performed these calculations explicitly, we expect other state-of-the-art quantum computers to deliver similar results.

In summary, we played two nonlocal games with 6 qubits using quantum computers on the cloud. Our work offers an important benchmark of quantum computers from different companies and technologies (superconducting qubits by IBM and trapped ions by IonQ and Honeywell). We, first, tested the minimal implementation of the triangle game of  Ref.~\cite{bravyi2018quantum} on 5 quantum computers on the cloud. We found that only one quantum computer demonstrated a winning probability of about 97\%, larger than the classical bound of 87.5\%, demonstrating a quantum advantage. In spite of our attempt to use best-practice optimization schemes compatible with this game, the other 4 computers had a winning probability lower than 87.5\% and, hence, failed to demonstrate quantum advantage. Interestingly, the Honeywell H1 is the most recently released quantum computer that was used in this work: its success to realize the triangle game may also reflect the steady improvement of quantum computers over time. To demonstrate quantum advantage in the other quantum computers, we, then, presented two nonlocal games inspired by known Bell inequalities \cite{guhne2005bell,cabello2008mermin,silman2008relation,cabello2020converting}. Specifically, we considered the optimal sum of products of stabilizers, $S_{\rm optimal}$, explicitly given by Eq.~(\ref{eq:optimal}). This operator is associated with a nonlocal game with a threshold to quantum advantage of $67\%$, which we cross by 20 standard deviations.

An interesting question for future study is the demonstration of quantum advantage with cluster states made of a larger number of qubits, $n>6$. For the triangle game of Ref.~\cite{bravyi2018quantum}, the number of circuits (8)  and the threshold to quantum advantage ($87.5\%$) are independent on $n$. Because the winning conditions of the game involve products of $n$ qubits, the error from each qubit sum up and it is effectively harder to show quantum advantage with increasing $n$. In contrast, the approach based on Bell inequalities can deliver a quantum advantage at a fixed error per qubit. Although $S_{\rm optimal}$ is known only for $n \le 6$ \cite{cabello2008mermin}, we can derive an upper bound to its threshold by splitting the cluster state in small parts and taking the product of  $S_{\rm optimal}$ of each part. This analysis suggests that the ratio between the maximal LHV value of $S_{\rm optimal}$ and its quantum value should decrease exponentially with $n$. Hence, the classical winning probability decreases with $n$, and can balance the effective increase of qubit errors. The price to pay is that the number of elements in $S_{\rm optimal}$ increases exponentially with $n$, hence requiring a prohibitive number of measurements. The optimization of the winning probability versus the number of measurements is an open question that needs to be addressed theoretically.

{\bf Acknowledgments} We acknowledge useful discussions with Tomer Simon, Eyal Malach (Microsoft Azure Quantum), and Eli Arbel, Haggai Landa (IBM Research Haifa). We acknowledge the use of IBM Quantum services for this work and we thank Eli Arbel for generating the results on the ibm\_boeblingen device. The views expressed are those of the authors, and do not reflect the official policy or position of IBM or the IBM Quantum team. We thank Brian Neyenhuis for generating the results on the Honeywell H1 device. We acknowledge the technical support of Eyal Estrin (Inter University Computation Center) and Tomer Ofir (Amazon Web Services).  This work was supported by the Israel Science Foundation, grants number 151/19 and 154/19, the AWS Cloud Credit for Research Program, and Microsoft Quantum Azure Israel.

\end{document}